\documentclass{aa}  

\usepackage{graphicx}
\usepackage{txfonts}
\usepackage{hyperref}
\usepackage[squaren]{SIunits}

\newcommand{\Msun}{{\rm M}_{\sun}}				

\title{Masses of double neutron star mergers}
\author{Matthias U. Kruckow}
\institute{Yunnan Observatories, Chinese Academy of Sciences, Kunming 650216, China\\\email{mkruckow@ynao.ac.cn}}
\date{Received ?? ??, 2020; accepted ?? ??, ????}
\abstract
   {}
   {The mass discrepancy between the observed population of double neutron star binaries by radio pulsar observations and gravitational-wave observation requires an explanation.}
   {Binary population synthesis calculations are performed, and their results are compared with the radio and the gravitational-wave observations simultaneously.}
   {Simulations of binary evolution are used to link different observations of double neutron star binaries with each other. The progenitor of GW190425 is investigated in more detail. A distribution of masses and merger times of the possible progenitors is presented.}
   {A mass discrepancy between the radio pulsars in the Milky Way with another neutron star companion and the inferred masses from gravitational-wave observations of those kind of merging systems is naturally found in binary evolution.}

\begin{document} 

\maketitle
%

\section{Introduction}\label{sec:Introduction}
After GW170817 \citep{aaa+17c}, a second candidate of a double neutron star (DNS) merger is recently reported by \citet{aaa+20}, namely GW190425. This event relates to a merging binary significantly more massive then the well known GW170817. When compared to the population of DNSs in the Milky Way, observed by radio pulses of at least one of the neutron stars (NSs), GW190425 is more massive then expected. The population of DNSs from pulsar observations has binary masses between $2.5$ and $\unit{2.9}{\Msun}$, while the underlying population may spans a range of $2.2$ to $\unit{3.2}{\Msun}$ \citep{fzt19}. The analysis of GW190425 shows that this event most probably originates from a merger with a binary mass of $\unit{3.4^{+0.3}_{-0.1}}{\Msun}$ \citep{aaa+20}.

During the binary evolution of systems forming DNSs several processes strongly influence the population and their properties \citep[e.g.][]{tkf+17,vns+18,am19,csh+19}. One important phase in the evolution is the supernova (SN), where the NSs form. In this phase, a kick imparted on the newborn NS changes the orbit. This can result in a widening and/or a more eccentric orbit. Merging systems need to be tight enough for efficient gravitational wave (GW) radiation. On the other hand, to observe one NS of a DNS as a radio pulsar only requires the binary to keep bound. Thus, high kicks may leave a system observable in radio, but remove it from being a merging candidate. Also, if a system evolves through a common envelope (CE) phase and would have been in the mass range to become a DNS it might merge early. On the other hand such a phase helps to tighten the DNS progenitors. Hence, a comparison between models and observations requires more care to be able to link the different observations with each other.

The physics included in the simulations is summarized in Sec.~\ref{sec:Methods}. Sec.~\ref{sec:Results} shows the outcomes of the binary population synthesis calculation related to the radio and GW observations. After presenting the possible progenitors of GW190425 a short discussion about the results follows in Sec.~\ref{sec:Discussions}. Finally, the conclusions are drawn.

\section{Methods}\label{sec:Methods}
\begin{table*}
 \centering
 \caption{\label{tab:parameters}Initial values and settings of key input physics parameters for the default and optimistic simulation, taken from tables 2 and 8 of \citet{ktl+18}. The simulation at lower metallicity used the same parameters as the default simulation at $Z=0.0047$.}
 \begin{tabular}{lll}
  \hline
  name & default simulation & optimistic simulation\\
  \hline
  number of simulated binaries, $N$ & $10^{9}$ & $10^{9}$\\
  initial mass function parameter, $\alpha_{\rm IMF}$ & $\enspace2.7$ \citep{sal55,sca86} & $\enspace2.3$ \citep{kro08}\\
  primary mass, $m^{\rm p}_{\rm ZAMS}$ & $\in\unit{[4:100]}{\Msun}$ & $\in\unit{[4:100]}{\Msun}$\\
  mass ratio distribution & \citet{kui35} & \citet{kui35}\\
  secondary mass, $m^{\rm s}_{\rm ZAMS}$ & $\in\unit{[1:100]}{\Msun}$ & $\in\unit{[1:100]}{\Msun}$\\
  period distribution & \citet{opi24,abt83} & \citet{opi24,abt83}\\
  semi-major axis, $a$ & $\in\unit{[2:10\,000]}{{\rm R}_{\sun}}$ & $\in\unit{[2:10\,000]}{{\rm R}_{\sun}}$\\
  eccentricity, $e$ & $\enspace0$ (initially circular orbit) & $\enspace0$ (initially circular orbit)\\
  metallicity, $Z$ & $\enspace0.0088$ (Milky Way-like) & $\enspace0.0088$ (Milky Way-like)\\
  rotation, $v_{\rm rot}$ & $\enspace\unit{0}{\kilo\meter\usk\reciprocal\second}$ (non-rotating stars) & $\enspace\unit{0}{\kilo\meter\usk\reciprocal\second}$ (non-rotating stars)\\
  \hline
  \multicolumn{3}{l}{during stable Roche-lobe overflow \citep{spv97}:}\\
  ~~wind mass loss, $\alpha_{\rm RLO}$ & $\enspace0.2$ & $\enspace0.15$\\
  ~~minimum mass ejection by accretor, $\beta_{\rm min}$ & $\enspace0.75$ & $\enspace0.5$\\
  ~~circumbinary torus mass transfer, $\delta_{\rm RLO}$ & $\enspace0$ & $\enspace0$\\
  ~~circumbinary torus size, $\gamma$ & $\enspace2$ & $\enspace2$\\
  \multicolumn{3}{l}{during CE:}\\
  ~~CE efficiency parameter, $\alpha_{\rm CE}$ & $\enspace0.5$ & $\enspace0.8$\\
  ~~fraction of internal energy, $\alpha_{\rm th}$ & $\enspace0.5$ & $\enspace0.3$\\
  critical mass ratio for mass transfer stability, $q_{\rm limit}$ & $\enspace2.5$ & $\enspace2.5$\\
  \hline
 \end{tabular}
\end{table*}

For this paper binary population synthesis simulations are performed, using the \textsc{ComBinE} code \citep[see][for details]{ktl+18}. The most crucial parts of the evolution towards a merger of two NSs are the CE evolution and the SN treatment.

\textsc{ComBinE} interpolates grids of stellar evolution to evolve binaries from the zero-age main sequence. First, the binaries are created from given initial distributions in the specified range of stellar masses, semi-major axis and eccentricities, see first half of Tab.~\ref{tab:parameters}. Additionally, some other parameters like metallicity are specified for each simulation. Those binaries are evolved taking the effects of nuclear burning, wind mass loss, tidal effects (mainly circularisation), and other effects of single star evolution into account. It is always checked whether there are binary interactions, e.g. Roche-lobe overflow, occurring. The parameters of mass transfer are summarized in the second half of Tab.~\ref{tab:parameters} and their detailed prescription can be found in \citet{ktl+18}, which mainly follows \citet{spv97} in the case of stable mass transfer. If mass transfer is initiated this could proceed in a stable way or lead to a CE phase.

During CE calculation the energy budget or $(\alpha,\lambda)$-formalism \citep{web84,dek90} is considered. Here \textsc{ComBinE} uses self consistent values of the envelope binding parameter $\lambda$, cf. equations (10) and (11) in \citet{ktl+18}, and allows taking thermal and recombination energy into account, too. This results in two free parameters $\alpha_\mathrm{CE}$ and $\alpha_\mathrm{th}$. $\alpha_\mathrm{CE}$ is the efficiency of converting orbital energy into kinetic energy of the envelope to eject it. The amount of usable thermal and recombination energy reflected in $\alpha_\mathrm{th}$ and therefore modifies the effective $\lambda$. In this way, both efficiency parameters are restricted by energy conservation to $0\le\alpha\le1$.

At the end of massive star evolution, NSs usually form in a core-collapse or electron capture SN. During this SN the newly formed NS may receives a kick due to asymmetries in the explosion. Depending on the kind of SN and the material in the envelope different kick distributions are assumed, cf. table~1 in \citet{ktl+18}. Some envelope material may get stripped previously to the explosion. This treatment includes the consideration of an ultra-stripped SN \citep{tlm+13,tlp15,sys+15,mmt+17}. The kick on the newly formed NS results in a change of the orbit of the binary \citep{tt98,ktl+18}. As a result the time until a merger due to GW radiation may change significantly. The strong dependence of the merging delay time, the time GW radiation takes to merge the binary, on the semi-major axis and the eccentricity \citep{pet64} plays a key role when comparing binaries of a certain mass but different orbital parameters.

\section{Results}\label{sec:Results}
\begin{figure*}
 \centering
 \includegraphics[width=0.495\textwidth]{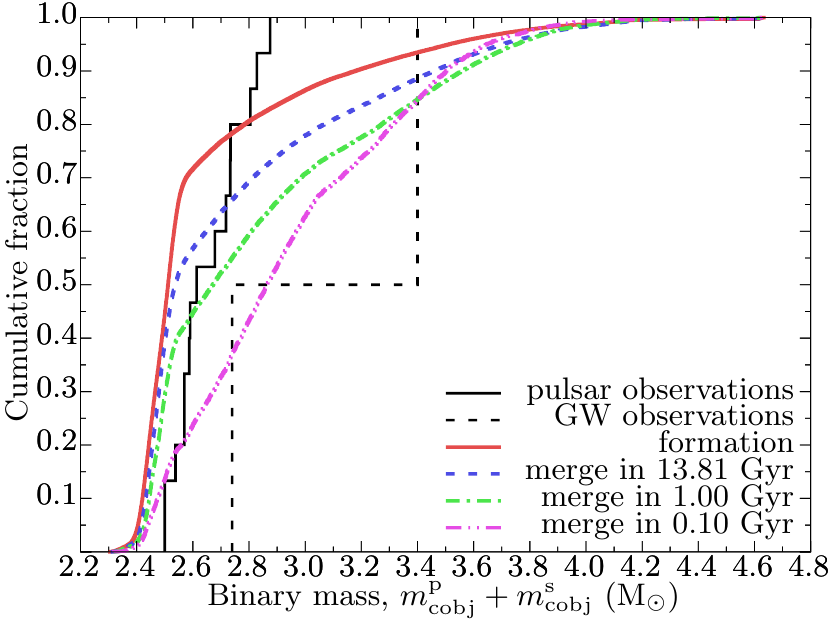}
 \includegraphics[width=0.495\textwidth]{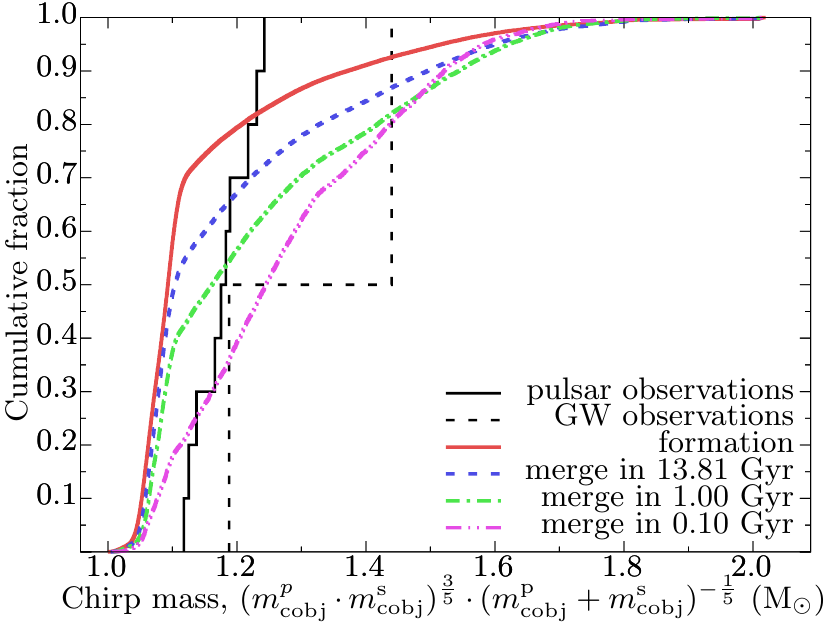}
 \includegraphics[width=0.495\textwidth]{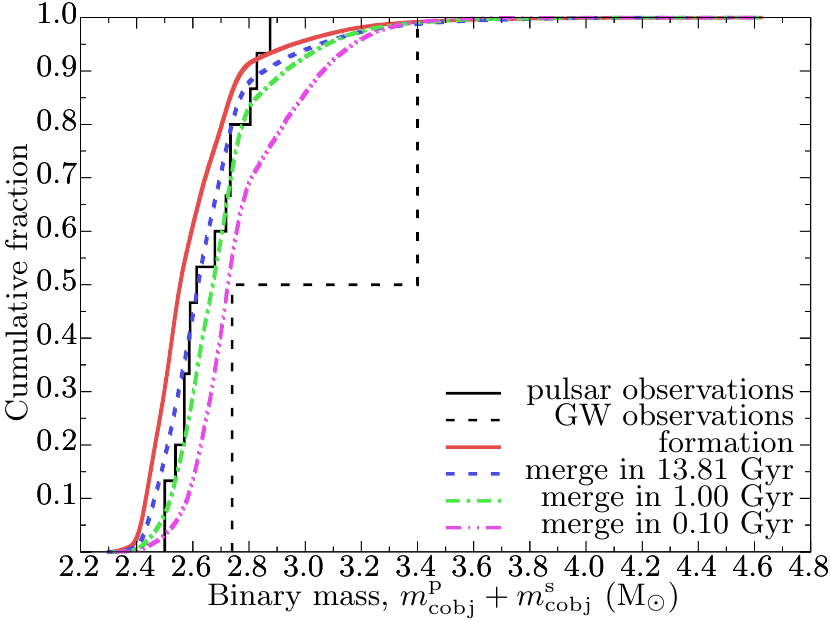}
 \includegraphics[width=0.495\textwidth]{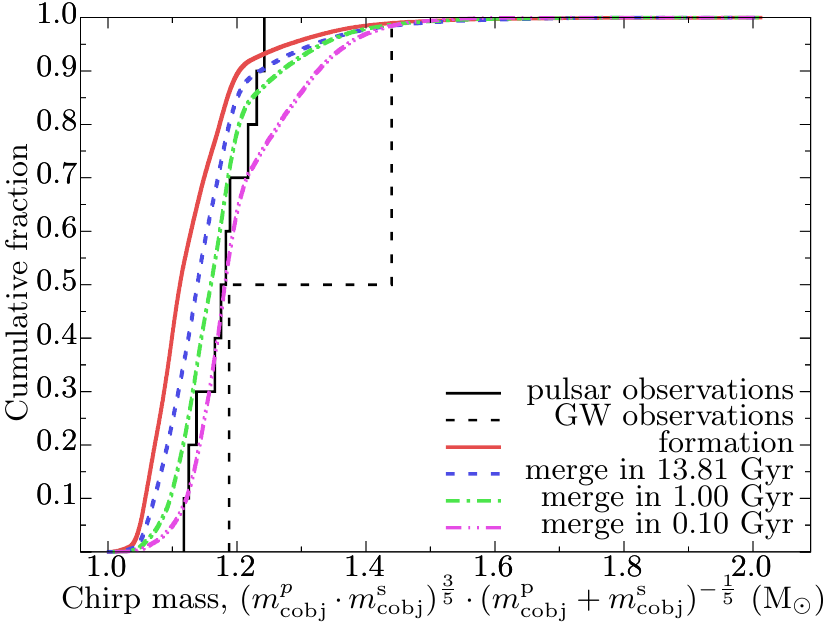}
 \caption{\label{fig:masses}The cumulative distribution of the binary masses (left) and the chirp masses (right). The top and bottom panels show the results of the default and optimistic simulation setup, respectively. The solid, red lines represent all formed double neutron star binaries in the simulation, which should be compared to the pulsar observations (solid, black). The dashed, black lines show the two reported merging events of neutron stars, which should be compared to the merging populations in the simulated data (dashed/blue, dash-dotted/green, dash-double dotted/purple).}
\end{figure*}
The simulations use the same parameters as described in \citet{ktl+18} and are again summarized in Tab.~\ref{tab:parameters}.

\subsection{Milky Way metallicity}
\begin{table}
 \centering
 \caption{\label{tab:counts}Number of DNS in the simulations with the default, optimistic and lower metallicity (LMC) setup.}
 \begin{tabular}{lrrr}
  \hline
  selection & default & optimistic & LMC\\
  \hline
  all formed DNS & $81023$ & $136616$ & $6532$\\
  merge in $\unit{13.81}{\giga\mathrm{yr}}$ & $33138$ & $78071$ & $2852$\\
  merge in $\unit{\enspace1.00}{\giga\mathrm{yr}}$ & $16205$ & $44386$ & $2723$\\
  merge in $\unit{\enspace0.10}{\giga\mathrm{yr}}$ & $3961$ & $16292$ & $2007$\\
  \hline
 \end{tabular}
\end{table}
Two simulations are performed at Milky Way metallicity ($Z=0.0088$). First, a default simulation, like in \citet{ktl+18}, is done. Second, an optimistic simulation with the aim of producing more DNS is calculated. The optimistic setup differs from the default one by using (i) a less steep initial mass function, (ii) a lower mass loss during stable mass transfer, (iii) a more efficient CE ejection, but using a bit less thermal and recombination energy there, cf. Tab~\ref{tab:parameters}.

Fig.~\ref{fig:masses} shows the cumulative distributions resulting from the simulations for the binary mass, $M=m_\mathrm{cobj}^\mathrm{p}+m_\mathrm{cobj}^\mathrm{s}$, and the chirp mass, $\mathcal{M}=(m_\mathrm{cobj}^\mathrm{p}\cdot m_\mathrm{cobj}^\mathrm{s})^\frac{3}{5}\cdot M^{-\frac{1}{5}}$, where $m_\mathrm{cobj}^\mathrm{p}$ and $m_\mathrm{cobj}^\mathrm{s}$ are the primary's and secondary's compact object mass, respectively. The simulated systems of all formed DNSs are subdivided into three groups depending on their delay time between star formation and the GW driven merging event. The number of DNSs in the simulations are summarized in Tab.~\ref{tab:counts}. Additionally, the observations from radio pulsars in the Milky Way \citep{clm+04,ksm+06,cks+07,wnt10,fst14,fsk+14,srm+15,vks+15,msf+15,msf+17,tkf+17,cck+18,fer18,lsk+18,sfc+18} and the two reported GW mergers \citep{aaa+17c,aaa+20} are put there for comparison. The masses obtained from observations are summarized in Tab.~\ref{tab:obs}.

Here it is important to note that the pulsar population, observed in the Galactic Disk, should be compared with all DNSs formed in the simulations, while the GW events should only resemble the simulated population which merges at least within a Hubble time after star formation. The populations which merge faster tend to consist of more massive NSs leading to a larger binary mass and chirp mass. While the default simulations better match the GW events (upper panels), the pulsar observations are better represented with the optimistic setup (lower panels). In both simulations the population of merging DNSs contains fewer low mass systems. The population merging within $\unit{0.1}{\giga{\rm yr}}$ after star formation contains the least less massive binaries. This can be seen in both mass distributions in Fig.~\ref{fig:masses} It should be noted, in both observational data sets, but especially for the GW events, the low number of observed systems may not represent the true population in the Universe.

\subsection{Lower metallicity}
\begin{figure*}
 \centering
 \includegraphics[width=0.495\textwidth]{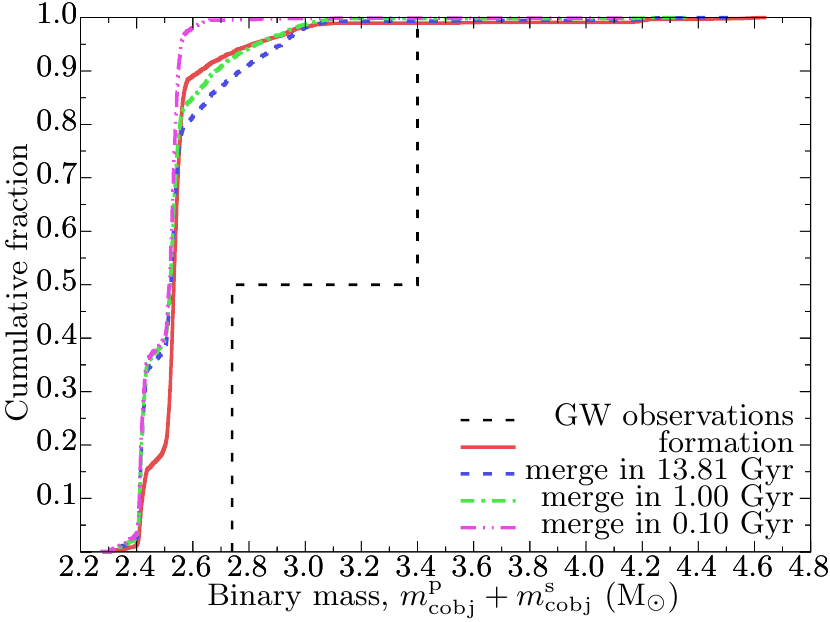}
 \includegraphics[width=0.495\textwidth]{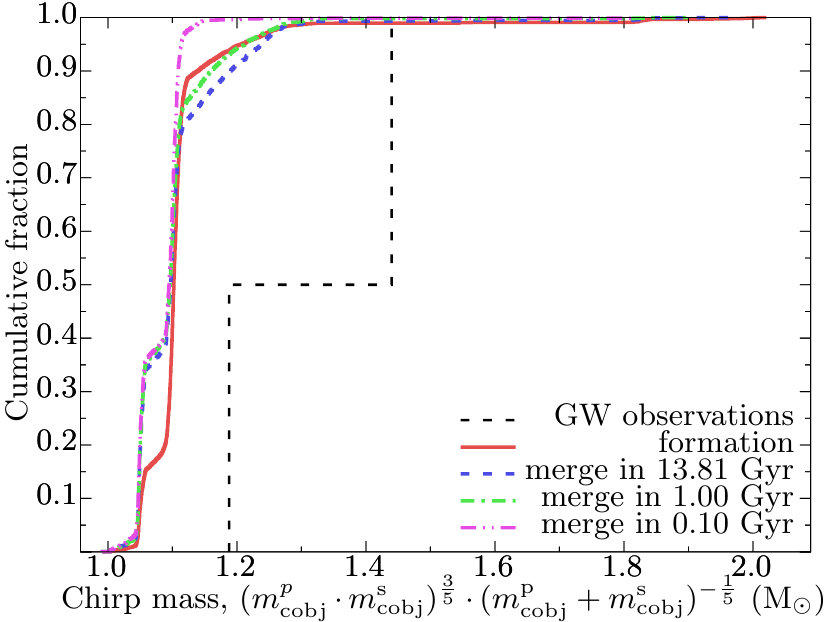}
 \caption{\label{fig:masses_LMC}Similar to the upper panels of Fig.~\ref{fig:masses} but for lower metallicity.}
\end{figure*}
At a lower metallicity, here close to that of the Large Magellanic Cloud ($Z=0.0047$), DNS binaries seem to favour smaller masses, see Fig.~\ref{fig:masses_LMC}. This even holds for the systems that merge fast enough to be observed as GW events. Here it turns out that most systems formed with low mass would merge in such a low metallicity environment. But one needs to be careful when comparing these results to the GW events as they have no information about the metallicity as long as there is no electromagnetic counterpart detected to identify the host galaxy. In the case of GW170817 the host galaxy has a metallicity comparable to that of the Milky Way, while in the case of GW190425 no host galaxy is identified. Furthermore, the different behaviour at lower metallicity requires more investigation in the future.

\subsection{GW190425}
\begin{figure}
 \centering
 \includegraphics[width=\columnwidth]{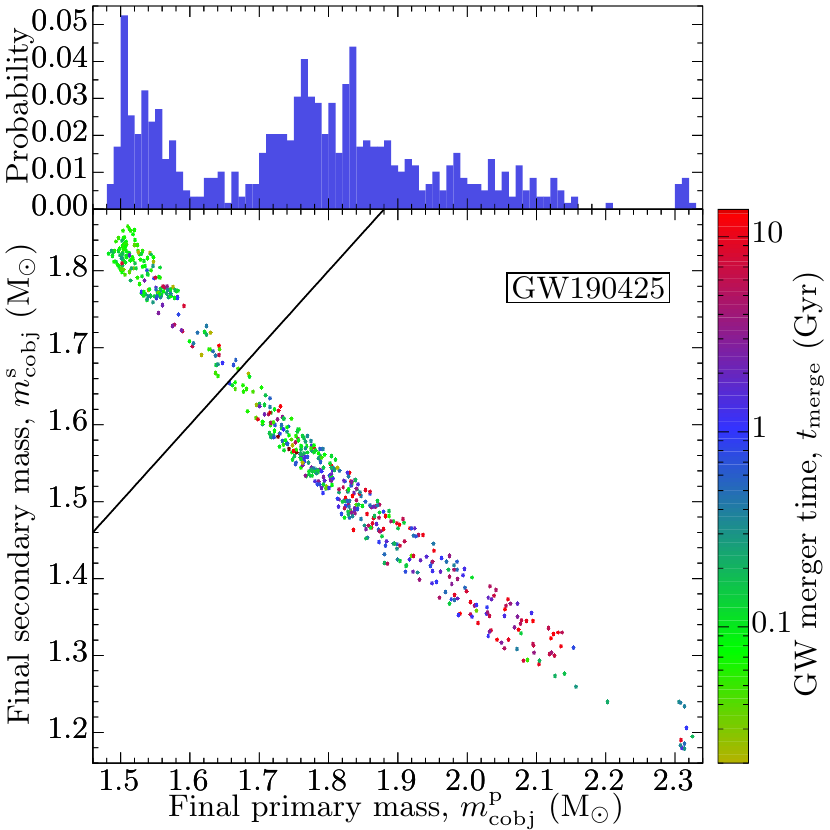}
 \caption{\label{fig:GW190425}Progenitor masses of GW190425 at Milky Way like metallicity in the default simulation. The color indicates the time until the GW merger. The Models are selected according to the reported binary mass of $\unit{3.4^{+0.3}_{-0.1}}{\Msun}$ and chirp mass of $\unit{1.44^{+0.02}_{-0.02}}{\Msun}$, where errors mark the 90 per cent confidence interval \citep{aaa+20}. The solid black line marks the binaries with equal primary and secondary mass. On the top are the primary masses collapsed to a probability distribution.}
\end{figure}
The recently reported event GW190425 is likely to be a DNS merger, but the involvement of a black hole cannot be rules out \citep{aaa+20,kfh+20}. The high mass of the system is thought to pose a problem for the formation of such a system, and thought to be inconsistent with mass distribution of the known radio pulsars in the Milky Way \citep{aaa+20}. Binary population synthesis calculations show that the formation of such a system at Milky Way-like metallicity is possible, see Fig.~\ref{fig:GW190425}. Furthermore, the larger error bars compared to GW170817 provide weaker constrains to the binary and chirp mass, the best constrained masses of a GW event.

The simulations cover progenitors with all kinds of merger delay times. There are two main populations. First, a population with the initially more massive star, here referred to as the primary, becoming a NS with a mass between $1.5$ and $\unit{1.6}{\Msun}$. In this case the star with a lower zero-age main sequence mass gains mass during mass transfer episodes and finally becomes the more massive NS. This population prefers short delay times until merger. Second, a population with the primary's mass mainly lying in between $1.7$ and $\unit{1.9}{\Msun}$. Here the primary is the more massive binary component. This population extends to masses of more than $\unit{2.1}{\Msun}$. The more massive primaries prefer longer delay times, getting closer to the age of the Universe. The GW observations are incapable of differentiating between the two populations as these observations do not contain information about the formation order of the two NSs. A third, small population exceeds $\unit{2.3}{\Msun}$ for the primary NS. In the simulations presented here, the spin evolution of NSs is not followed. \citet{aaa+20} predicts that the mass interval of the more massive component covers masses above $\unit{2.5}{\Msun}$ for their high-spin prior. It is unclear whether such massive NSs exist \citep[e.g.][]{of16} and how they form.

The main formation channel for the GW190425 progenitors in the simulations contains a stable case B or case C mass transfer, in which the mass transfer initiates after the primary leaves the main sequence, onto its companion. After the primary explodes to form the first NS, the secondary evolves to fill its own Roche lobe. This leads to a CE phase reducing the separation significantly. Before the CE, the binary needs to have a large separation to have the secondary star to become a giant as donor star with a less tightly bound envelope. Only progenitor systems with longer delay times do not experience another mass transfer in the term of case BB, the mass transfer from a helium star after core helium exhaustion, leading to an ultra-stripped SN \citep{tlm+13}. This formation channel including the case BB mass transfer is similar to the formation channel proposed by \citet{rfs+20}. In the default simulation the zero-age main sequence mass of the primary ranges from $16$ to $\unit{30}{\Msun}$, while the initial mass of the secondary is between $10$ and $\unit{16}{\Msun}$. The initial binary semi-major axis covers a range from $400$ to $\unit{2000}{\mathrm{R}_\sun}$, hence being wide enough to initiate the first mass transfer only when the primary becomes a giant.

\section{Discussions}\label{sec:Discussions}
At Milky Way metallicity merging DNSs tend to have a larger binary and chirp mass than the entire population of DNSs formed. The set of DNSs which is only part of the formed but not merging population contains the systems having a large separation. Those binaries require smaller kicks from the SN in order to remain bound. Electron capture SNe are expected to have smaller kicks, as they eject less material. Thus, NSs formed in an electron capture SN can more easily create wider DNSs. As the merging population did not include those systems, the relative abundance of more massive DNSs is larger for this population. This coincides with the fact that many DNSs from radio pulsar observations in the Milky Way have long merger delay times and would not merge within a Hubble time via GW radiation \citep{tkf+17}. It should be noted that there is still a mass discrepancy between the most common population in the simulation, mainly formed via electron capture SNe, and the pulsar observations \citep{ktl+18}. Beside this small mass discrepancy the population from the optimistic simulation is similar to the distribution found in \citet{fzt19}. The only other difference is that here the most massive DNSs are a bit more massive. Another possibility of enriching the massive binary content of the merging population is a dynamical tightening in star clusters \citep[e.g.][]{yfk+20}.

The situation at lower metallicity differs from the simulations at Milky Way-like metallicity, cf. Figs~\ref{fig:masses} and \ref{fig:masses_LMC}. The simulations at lower metallicity should be taken with a grain of salt as there are no strong constrains provided by observations of binaries containing NSs at lower metallicity/outside the Milky Way. There are no radio pulsar observations of DNSs or detailed observations of X-ray binaries as progenitors of those. Hence, the populations in the Milky Way and other galaxies may differ for other reasons \citep{pan18}.

The progenitor of GW190425, see Fig.~\ref{fig:GW190425}, shows possible component masses up to $\unit{2.3}{\Msun}$, which is consistent with the recent results of \citet{wsj20} for NS masses from helium stars at solar metallicity. Other investigations of double neutron star mergers \citep[e.g.][]{mgs+19} suggest individual component masses are just below $\unit{2.0}{\Msun}$. Masses up to $\unit{2.7}{\Msun}$, as suggested by the high-spin analysis of GW190425 \citep{aaa+20} are not obtained by calculations of binary evolution. This may indicate a less massive fast spinning or a slow spinning NS as the more massive component in the merging event. Furthermore, a possible black hole origin cannot be ruled out. However it should be noted that such a less massive black hole would challenge isolated binary evolution as well.

\section{Conclusions}\label{sec:Conclusions}
Using the default setup, the population of DNS merging within $\unit{0.1}{\giga\mathrm{yr}}$ coincides well with GW170817 and GW190425. If these two events are representative of further GW detections of merging DNS, they match the simulations. At the same time the simulations produce less massive systems in wider orbits. Those are more consistent with the radio pulsar population observed in the Milky Way. Because of the low number of observations the mass distribution of merging DNSs is not finalised yet. But it is shown that the statement that the pulsar population and the DNS merger population are not consistent cannot be drawn.

The recently reported event GW190425 has possible progenitors from the current understanding of binary evolution. Depending on the detailed component masses the delay time of the merger differs significantly. The simulations cannot reproduce the high-spin models with a very high mass component. Due to its host galaxy have not been identified, no information about the metallicity of the progenitor and its possible delay time since star formation can be used to further narrow the mass estimate of the event.

\begin{acknowledgements}
I would like to thank Hailiang Chen for some feedback to this paper. This work is partly supported by Grant No 11521303 and 11733008 of the Natural Science Foundation of China.
\end{acknowledgements}

\bibliographystyle{aa} 
\bibliography{kruckow_refs} 

\appendix
\section{Observational data}
\begin{table*}
 \centering
 \caption{\label{tab:obs}Summary of the GW events, first table, and the pulsar observations, second table.}
 \begin{tabular}{llll}
  \hline
  event name & binary mass & chirp mass & reference\\
  \hline
  GW170817 & $\unit{2.74_{-0.01}^{+0.04}}{\Msun}$ & $\unit{1.188_{-0.002}^{+0.004}}{\Msun}$ & \citet{aaa+17c}\\
  GW190425 & $\unit{3.4\phantom{0}_{-0.1\phantom{0}}^{+0.3\phantom{0}}}{\Msun}$ & $\unit{1.44\phantom{0}_{-0.02\phantom{0}}^{+0.02\phantom{0}}}{\Msun}$ & \citet{aaa+20}\\
  \hline
 \end{tabular}
 \\[\baselineskip]
 \begin{tabular}{lllll}
  \hline
  pulsar name & pulsar mass & companion mass & binary mass & reference\\
  \hline
  J0453+1559 & $\unit{1.559\phantom{0}\pm0.005\phantom{0}}{\Msun}$ & $\unit{1.174\phantom{0}\pm0.004\phantom{0}}{\Msun}$ & $\unit{2.734\phantom{000}\pm0.004\phantom{000}}{\Msun}$ & \citet{msf+15}\\
  J0509+3801 & $\unit{1.34\phantom{00}\pm0.08\phantom{00}}{\Msun}$ & $\unit{1.46\phantom{00}\pm0.08\phantom{00}}{\Msun}$ & $\unit{2.805\phantom{000}\pm0.003\phantom{000}}{\Msun}$ & \citet{lsk+18}\\
  J0737-3039 & $\unit{1.3381\pm0.0007}{\Msun}$ & $\unit{1.2489\pm0.0007}{\Msun}$ & $\unit{2.58708\phantom{0}\pm0.00016\phantom{0}}{\Msun}$ & \citet{ksm+06}\\
  J1411+2551 & $\unit{<1.62}{\Msun}$ & $\unit{>0.92}{\Msun}$ & $\unit{2.538\phantom{000}\pm0.022\phantom{000}}{\Msun}$ & \citet{msf+17}\\
  J1518+4904 & $\unit{1.41\phantom{00\pm0.0000}}{\Msun}$ & $\unit{1.31\phantom{00\pm0.0000}}{\Msun}$ & $\unit{2.718\phantom{000\pm0.000000}}{\Msun}$ & \citet{tkf+17}\\
  B1534+12 & $\unit{1.3330\pm0.0002}{\Msun}$ & $\unit{1.3455\pm0.0002}{\Msun}$ & $\unit{2.678463\pm0.000004}{\Msun}$ & \citet{fst14}\\
  J1756-2251 & $\unit{1.341\phantom{0}\pm0.007\phantom{0}}{\Msun}$ & $\unit{1.230\phantom{0}\pm0.007\phantom{0}}{\Msun}$ & $\unit{2.56999\phantom{0}\pm0.00006\phantom{0}}{\Msun}$ & \citet{fsk+14}\\
  J1757-1854 & $\unit{1.3384\pm0.0009}{\Msun}$ & $\unit{1.3946\pm0.0009}{\Msun}$ & $\unit{2.73295\phantom{0}\pm0.00009\phantom{0}}{\Msun}$ & \citet{cck+18}\\
  J1811-1736 & $\unit{<1.64}{\Msun}$ & $\unit{>0.93}{\Msun}$ & $\unit{2.57\phantom{0000}\pm0.10\phantom{0000}}{\Msun}$ & \citet{cks+07}\\
  J1829+2456 & $\unit{<1.38}{\Msun}$ & $\unit{>1.22}{\Msun}$ & $\unit{2.5\phantom{00000}\pm0.2\phantom{00000}}{\Msun}$ & \citet{clm+04}\\
  J1906+0746 & $\unit{1.291\phantom{0}\pm0.011\phantom{0}}{\Msun}$ & $\unit{1.322\phantom{0}\pm0.011\phantom{0}}{\Msun}$ & $\unit{2.6134\phantom{00}\pm0.0003\phantom{00}}{\Msun}$ & \citet{vks+15}\\
  J1913+1102 & $\unit{1.65\phantom{00}\pm0.05\phantom{00}}{\Msun}$ & $\unit{1.24\phantom{00}\pm0.05\phantom{00}}{\Msun}$ & $\unit{2.875\phantom{000}\pm0.14\phantom{0000}}{\Msun}$ & {\small\citet{fer18}}\\
  B1913+16 & $\unit{1.4398\pm0.0002}{\Msun}$ & $\unit{1.3886\pm0.0002}{\Msun}$ & $\unit{2.828378\pm0.000007}{\Msun}$ & \citet{wnt10}\\
  J1930-1852 & $\unit{<1.32}{\Msun}$ & $\unit{>1.30}{\Msun}$ & $\unit{2.59\phantom{0000}\pm0.04\phantom{0000}}{\Msun}$ & \citet{srm+15}\\
  J1946+2052 & - & - & $\unit{2.50\phantom{0000}\pm0.04\phantom{0000}}{\Msun}$ & \citet{sfc+18}\\
  \hline
 \end{tabular}
\end{table*}
Information about DNS are obtained from pulsar observation in the Milky Way and inferred from GW merging events. All the observational data on the masses of DNS can be found in Tab.~\ref{tab:obs}. The chirp masses for the DNS in the Milky Way are calculated from the individual masses in the ten cases, where this data is available.

The pulsar observations are limited to the Milky Way, while the GW observations lacking of precise sky localisation. The pulsar observations are incomplete for binaries with very long orbital periods ($\unit{>1}{\rm yr}$). Additionally the discoveries of pulsars in a binary is more likely for systems with lower acceleration, hence not eccentric and/or not too tight. On the other hand, the GW observations provide no information about the local environment of an event as long as no electromagnetic counterpart is identified. GW events with larger masses have a larger GW amplitude, which helps to identify them close to the detection limit.

\end{document}